\documentclass[letter]{ptptex}

\usepackage{amsmath,amssymb}
\usepackage{graphicx}

\notypesetlogo

\newcommand{\bra}[1]{{\!}\left\langle{}{ #1 }{}\right|{}}
\newcommand{\ket}[1]{{}\left|{}{ #1}{}\right\rangle{\!}}
\newcommand{\bracket}[2]{{\!}\left\langle{}{#1}\left.\!\!\vphantom{#2}%
\vphantom{#1}{}\right|{}\!{#2}{}\right\rangle{}\!}
\newcommand{\ev}[1]{\left\langle{}{#1}\right\rangle{}}

\renewcommand{\vec}[1]{\mathbf #1}
\newcommand{\nn}{\nonumber}
\newcommand{\sla}[1]{{\not \! \! \: {#1}}}

\title{Structures of Expectation Values of Flavor Neutrino Charges with 
Respect to Neutrino-Source Hadrons
}
\author{Kanji Fujii and Takashi Shimomura}

\inst{Department of Physics, Faculty of Sciences, Hokkaido University, Sapporo 060-0810, Japan}
\date{\today}

\abst{
In the framework of quantum field theory, we examine what physical implications
are included in the expectation values of the flavor neutrino charges at the time $x^0$ with
respesct to the state $\ket{\Psi(x^0)}$, which is taken so as to coincide with the
 neutrino source state such as a charged pion at the time $x^0_I$ ($< x^0$).
}

\begin{document}
\maketitle

\noindent {\it Purpse} --- In the quantum field theory\cite{rf:1}, the expectation value of a physical 
observable $F(x)$ at a space-time point $x = (\vec{x},x^0)$ with respect
to the state $\ket{\Psi(x^0)}$ is given by, in the interaction
representation, 
\begin{align}
\bra{\Psi(x^0)}F(x)\ket{\Psi(x^0)}=\bra{\Psi(x^0_I)}S^{-1}(x^0,x^0_I)F(x)S(x^0,x^0_I)\ket{\Psi(x^0_I)},\\
S(x^0,x^0_I)=\sum_{m=0}(-i)^m\int^{x^0}_{x^0_I}d^4y_1\int^{y^0_1}_{x^0_I}d^4y_2\cdots\int^{y^0_{m-1}}_{x^0_I}d^4y_{m}
 H_{int}(y_1)\cdots H_{int}(y_m).
\end{align}

In the following we consider the expectation values of the flavor neutrino charges 
$N_\rho(x^0), \rho=$ the flavor suffix $(e, \mu, \cdots)$, with respect
to the state $\ket{\Psi(x^0)}$ which coincides with the neutrino-source
state $\ket{\Psi(x^0_I)}$ such as a charged pion at the time $x^0_I$ ($< x^0$), and point out that those 
expectation values lead to a possible way for treating unifiedly various weak decay 
probabilities of neutrino-source particles as well as the neutrino oscillation, and also give a 
suggestion how to construct the state of flavor-neutrino with momentum $\vec{k}$.

\noindent {\it Relevant quantities and relations} --- The total Lagrangian
$\mathcal{L}(x)=\mathcal{L}_0(x)+\mathcal{L}_{int}(x)$ related to 
neutrinos is taken to be 
\begin{align}
\mathcal{L}_0(x)&=-\bar{\nu}_F(x)[\sla{\partial}+M]\nu_F(x),\qquad
 M^\dagger=M,\\
\mathcal{L}_{int}(x)&=-[\bar{\nu}_F(x)J_F(x)+\bar{J}_F(x)\nu_F(x)]=-H_{int}(x);
\end{align}
here, $\nu_F(x)$ is a set of the flavor neutrino fields $\nu_\rho(x)$'s,
and a set of the mass eigenfields $\nu_j(x)$'s is defined by 
\begin{align}
\nu_F(x)=Z^{1/2}\nu_M(x),\ {Z^{1/2}}^\dagger M Z^{1/2}=M_{diag},\quad
 {Z^{1/2}}^\dagger Z^{1/2}=I,\ Z^{1/2}=[Z^{1/2}_{\rho j}];\\
\nu_F(x)=
\begin{bmatrix}
\nu_e(x)\\
\nu_\mu(x)\\
\vdots
\end{bmatrix},\quad
\nu_M(x)=
\begin{bmatrix}
\nu_1(x)\\
\nu_2(x)\\
\vdots
\end{bmatrix},\quad
\text{diag}(M_{diag})=(m_1,m_2,\cdots)\ ;
\end{align}
the source function $J_F(x)$ is for simplicity assumed to include no
neutrino fields, and $\mathcal{L}_{int}(x)$ is the local Fermi
interaction obtained in accodance with the electro-weak unified theory.
We employ Greek and Roman indices, $(\rho, \sigma, \lambda, \cdots)$ and
$(i,j,l,\cdots)$, to represent the flavor 
and the mass degrees of freedom, respectively.
 
It is noted that $F^H(x)\equiv S^{-1}(x^0,x^0_I)F(x)S(x^0,x^0_I)$
appearing in (1) is a quantity in Heisenberg 
representation, which is taken so as to coincide with the interaction representation at the 
time $x^0_I$. Hereafter we denote quantities in Heisenberg and the interaction representations 
by attaching and omitting the prefix $H$, respectively. We see that, due to (5),           
$(\sla{\partial}+M_{diag})\nu_M(x)=0$ leads to
$(\sla{\partial}+M)\nu_F(x)=0$ 
and $(\sla{\partial}+M)\nu_F^H(x)=-J^H_F(x)$. The integral of the latter is
easily expressed in the form of so-called
Yang-Feldman equation\cite{rf:2} when the field-mixing exists. Its implication in the neutrino problem will be discussed
elsewhere. In the following 
we consider directly the perturbative expansion of $S(x^0,x^0_I)$ with respect to the weak 
interaction 
\begin{align}
H_{int}(x)=\sum_\sigma \frac{G_F}{\sqrt{2}}i\bar{\nu}_\sigma(x)v^a
 l_\sigma(x) j^{had}_a(x)^\dagger+H.c.,
\end{align}
where $v^a=\gamma^a(1+\gamma_5)$; $G_F$ is Fermi coupling constant. We employ Kramers representation\cite{rf:3} 
of $\gamma^a$-matrices,$\overrightarrow{\gamma}=-\rho_y\otimes\overrightarrow{\sigma}$,
$\gamma^4=\rho_x\otimes I$, $\gamma_5=-\rho_z\otimes I$. The flavor
neutrino charge $N_\rho$,
the expectation value of which is to be examined, is defined by 
\begin{align}
N_\rho(x)\equiv :i\int d^3 x j^4_{(\rho)}(\vec{x},x^0):,
\end{align}
where the 4-vector current of $\nu_\rho$-field is given by 
$j^a_{(\rho)}(x)=-i\bar{\nu}_{\rho}(x)\gamma^a\nu_\rho(x)$.
The symbol $:\ :$ represents the normal product with respect to the momentum-helicity 
creation and annihilation operstors of the $\nu_M(x)$ field, which is expanded as 
\begin{align}
\nu_i(x)=\sum_{\vec{k},r}\frac{1}{\sqrt{V}}
 \left[\alpha_j(kr)u_j(kr)e^{i(kx)}+\beta_j^\dagger(kr)v_j(kr)e^{-i(kx)}\right];
\end{align}
here $(kx) = \vec{k}\cdot\vec{x} -\omega_j(k)x^0$ with
$\omega_j(k)=\sqrt{\vec{k}^2+m_j^2}$; 
"$r$" represents the helicity, $r = \uparrow\ \downarrow$; 
$(i\sla{k}+m_j)u_j(kr)=0,\ (-i\sla{k}+m_j)v_j(kr)=0$; 
$\alpha_j$, $\beta_j$ and their Hermitian conjugates satisfy
$\{\alpha_j(kr),\alpha^\dagger_i(k'r')\}=
\{\beta_j(kr),\beta^\dagger_i(k'r')\}=\delta_{ji}\delta_{rr'}\delta(\vec{k},\vec{k}')$,
others $= 0$; the concrete forms of $u_j(kr)$ and $v_j(kr)$ are given in Refs.\citen{rf:3,rf:4,rf:5}. 

\noindent {\it Expectation values of flavor neutrino charges} --- We investigate the structure of 
\begin{align}
\bra{A(x^0_I)}S^{-1}(x^0,x^0_I)N_\rho(x^0)S(x^0,x^0_I)\ket{A(x^0_I)}=
 \bra{A(x^0_I)}N_\rho^H(x^0)\ket{A(x^0_I)},
\end{align}
where $\ket{A(x^0_I)}$ $(x^0_I\leq x^0)$ is a hadronic state with no neutrinos and plays a role of a neutrino 
source, such as one-pion state.   For simplicity, we consider the case where an A-particle is a (pseudo-)scalar one with decay modes caused by (7).   The contributions of 
$G_F$ 0-th  and first orders are easily seen to vanish.   There are two kinds of $G_F^2$-order 
connected contributions, expressed as
\begin{align}
\bra{A(x^0_I)}&N_\rho^H(x^0)\ket{A(x^0_I)}_{con(I)}\nn\\
&\equiv 
 \bra{A(x^0_I)}\int^{x^0}_{x^0_I}d^4z \int^{x^0}_{x^0_I}d^4y
 H_{int}(z)N_\rho(x^0)H_{int}(y) \ket{A(x^0_I)}_{con},\\
\bra{A(x^0_I)}&N_\rho^H(x^0)\ket{A(x^0_I)}_{con(II)}
 \equiv 
 \bra{A(x^0_I)} i^2 \int^{x^0}_{x^0_I}d^4y\int^{y^0}_{x^0_I}d^4z\nn\\
&\times\left[H_{int}(z)H_{int}(y)N_\rho(x^0)+ N_\rho(x^0) H_{int}(y)H_{int}(z)\right]\ket{A(x^0_I)}_{con}.
\end{align}
We examine dominant contributions coming from diagrammatically lower
configurations in intermediate states.  Those contributions come from (11) and, in the 
case of a positively charged A-particle, are expressed as  
\begin{align}
&\ev{N_\rho(x^0);A^+(x^0_I)}_{con} \equiv \int d^3x
 \int^{x^0}_{x^0_I}d^4z \int^{x^0}_{x^0_I}d^4y \bra{A^+(x^0_I)}\nn\\
&\times\sum_{\lambda
 \sigma}\bar{J}_\lambda(z)\gamma^4\mathcal{S}_{\rho\lambda}^{(+)}(x-z)^\dagger
 \cdot \mathcal{S}_{\rho\sigma}^{(+)}(x-y)J_\sigma(y)\ket{A^+(x^0_I)}_{con}.
\end{align}
where $\bra{0}\nu_\rho(x)\bar{\nu}_\sigma(y)\ket{0}=-i\mathcal{S}^{(+)}_{\rho\sigma}(x-y)$.

The state $\ket{A^\dagger(x^0_I)}$ is now taken to be a plane-wave,
i.e. $\ket{A^\dagger(p,x^0_I)}=\alpha^\dagger_A(p)\ket{0}e^{iE_A(p)x^0_I}$
with $p^0=E_A(p)=\sqrt{\vec{p}+m_A^2}$,
$[\alpha_A(p),\alpha^\dagger_A(p')]=\delta(\vec{p},\vec{p}')$.
RHS of (13) includes a hadronic part, expressed as 
\begin{align}
&\bra{A^+(p,x^0_I)}j_a^{had}(z)\cdot
 j_b^{had}(y)^\dagger\ket{A^+(p,x^0_I)}
 =
 \bra{A^+(p,x^0_I)}j_a^{had}(z)\bigg\{\ket{0}\bra{0}\nn\\
&+\sum\ket{1\text{-particle}}\bra{1\text{-particle}}
 +\text{higher configurations}\bigg\}j^{had}_b(y)^\dagger\ket{A^+(p,x^0_I)}.
\end{align}
The intermediate vacuum contribution is written as  
\begin{align}
&\ev{N_\rho(x^0); A(x^0_I)}^{vac}_{con}=
 \left[i\frac{G_F}{\sqrt{2}}f_A\right]^2 
 \int^{x^0}_{x^0_I} dz^0 \int^{x^0}_{x^0_I} dy^0
 \frac{1}{2E_A(p)V}\sum_{\vec{q}}\sum_{\vec{k}}
 \delta(\vec{k}+\vec{q},\vec{p})\nn\\
&\times \sum_{j,i,\sigma} Z^{1/2}_{\sigma j} {Z^{1/2}_{\rho j}}^\ast
 Z^{1/2}_{\rho i} {Z^{1/2}_{\sigma i}}^\ast
 \sum_{s,r}\bar{v}_{l\sigma}(qs)\sla{p}(1+\gamma_5)
 u_j(kr)\rho_{ij}(k) \cdot \bar{u}_i(kr)\sla{p}(1+\gamma_5)v_{l\sigma}(qs)\nn \\
&\times e^{i(\omega_j-\omega_i)x^0}e^{i(-\omega_j-E_\sigma+E_A)z^0}
 e^{i(\omega_i+E_\sigma-E_A)y^0};
\end{align}
here, $f_A$ is the decay constant of $A^+$ , defined by   
\begin{align}
\bra{0}j^{had}_a(x)^+ \ket{A^+(p,x^0_I)}=\frac{1}{\sqrt{2E_A(p)V}}ip_a f_A
 e^{i(px)}\exp(iE_A(p)x^0_I);
\end{align}
$v_{l\sigma}(qs)$ is the momentum-helicity eigenfunction for $\bar{l}_\sigma$, satisfying
$(-i\sla{q}+m_\sigma)v_{l\sigma}(qs)=0$, $
q^0=E_\sigma(q)=\sqrt{\vec{q}^2+m_\sigma^2} $;
$u^\dagger_j(kr)u_i(ks)=\rho_{ji}(k)\delta_{rs}$, $\rho_{ji}(k)=\cos[(\chi_j-\chi_i)/2]$
 with $\cot\chi_j=|\vec{k}|/m_j$.\cite{rf:4,rf:5} 
Eq.(15) can be thought to lead to an oscillation formula.
Some simplification of (15) is obtained if we multiply by hand to RHS of (15) a damping 
factor due to the decay width $\Gamma_A(p)$ of A-particle, and perform 
the $z^0$- and $y^0$-integrations under the condition
$t\Gamma_A(p)\gg 1,\ t\equiv x^0-x^0_I$.  
Although the obtained expression includes the simple $t$-dependent factor
$exp[i(\omega_j(k)-\omega_i(k))t]$, we have to perform further the $\vec{k}$-integration; thus
the oscillation behavior is different from that of 
the well-known standard formula \cite{rf:6,rf:7} as well as of the expectation value of the flavor 
neutrino charge with respect to a flavor neutrino state\cite{rf:8}. Detailed analyses in the plane-
wave and the wave-packet $\ket{A(x^0_I)}$ cases will be done in a subsequent paper. ( Some 
considerations on the latter case have been done in Ref.9). )      

\noindent {\it Decay probabilities of neutrino source particles} --- 
We examine what results are obtained from $\ev{N_\rho(x^0); A^+(p,x^0_I)}$ for
sufficiently large $T=x^0-x^0_I$. We obtain from (15) $\ev{N_\rho(x^0); A^+(x^0_I)}^{vac}_{con}/T$
\begin{align}
\longrightarrow&
 \sum_{j,\sigma}|Z^{1/2}_{\rho j}|^2 |Z^{1/2}_{\sigma j}|^2 \frac{1}{(2\pi)^6}
 \int d^3k \int d^3q \frac{(2\pi)^4 \delta^4(k+q-p)}{2E_A(p)}
 \left[\frac{G_F}{\sqrt{2}}f_A\right]^2 \nn\\
&\times 
 \sum_{s,r} i\bar{v}_{l\sigma}(qs)\sla{p}(1+\gamma_5)
 u_j(kr)\cdot i\bar{u}_i(kr)\sla{p}(1+\gamma_5)v_{l\sigma}(qs)\nn \\
&={\sum_{j,\sigma}}'|Z^{1/2}_{\rho j}|^2 |Z^{1/2}_{\sigma j}|^2 P(A^+(p)\rightarrow \bar{l}_\sigma\nu_j);
\end{align}
here, $P(A^+(p)\rightarrow \bar{l}_\sigma\nu_j)=P(A^+(\vec{p}=0)\rightarrow
 \bar{l}_\sigma\nu_j)\cdot (m_A/E_A(p))$; the sum $\sum'_{j,\sigma}$ is
performed over $j$'s and $s$'s which are allowed under 4-momentum
conservation; the explicit form of the decay 
probability $P(A^+(\vec{p}=0)\rightarrow \bar{l}_\sigma\nu_j)$ calculated 
in the lowest order of the weak interaction  (7) is found e.g. in Refs.\ \citen{rf:5} and \citen{rf:10}.
 
By taking into account various intermediate states as shown in (14), we can generalize 
(17) to each weak semileptonic decay, $A^+\rightarrow \bar{l}_\sigma\nu_j
+\text{hadron(s)}$; then, by summing over all modes, we obtain in order
\begin{align}
&\left[\ev{N_\rho(x^0); A(x^0_I)}_{con}/T\right]_{T=x^0-x^0_I\rightarrow \infty}\quad
 (\equiv \ev{N_\rho; A^+(p),\infty})
 ={\sum_{j,\sigma}}'|Z^{1/2}_{\rho j}|^2 |Z^{1/2}_{\sigma j}|^2\nn\\
&\times\bigg[P(A^+(p)\rightarrow
 \bar{l}_\sigma\nu_j)+\sum_{\text{modes}}P(A^+(p)\rightarrow
 \bar{l}_\sigma\nu_j+\text{hadron(s)})\bigg], \\
&\sum_\rho \ev{N_\rho; A^+(p),\infty} \nn\\
&= {\sum_{j,\sigma}}'\big|Z^{1/2}_{\sigma j}\big|^2
\bigg[P(A^+(p)\rightarrow
 \bar{l}_\sigma\nu_j)+\sum_{\text{modes}}P(A^+(p)\rightarrow
 \bar{l}_\sigma\nu_j+\text{hadron(s)})\bigg].
\end{align}
Judging from the meaning of the expectation value now considering, LHS of (19) may be 
regarded as the total leptonic decay probability of $A^+$-particle accompanying a neutrino; in 
the case of $\pi^+$, its life-time $\tau(\pi^+(p))$ is given by
\begin{align}
\tau(\pi^+(p))^{-1}\simeq 
{\sum_{j,\sigma}}'\big|Z^{1/2}_{\sigma j}\big|^2
 P(\pi^+(p)\rightarrow \bar{l}_\sigma\nu_j)
 \simeq 
{\sum_j}'\big|Z^{1/2}_{\mu j}\big|^2
 P(\pi^+(p)\rightarrow \mu^+\nu_j).
\end{align}

It seems worthwhile to note the relation in the case of the expectation value of the 
number of charged lepton $l_\rho$. We obtain 
\begin{align}
&\left[\bra{A^+(p,x^0_I)}N^{H}_{l\rho}(x^0)\ket{A^+(p,x^0_I)}_{con}/T\right]_{T\rightarrow
 \infty}={\sum_j}'|Z^{1/2}_{\rho j}|^2\nn\\
&\times \left[P(A^+(p)\rightarrow
 \bar{l}_\rho\nu_j)+\sum_{\text{modes}}P(A^+(p)\rightarrow
 \bar{l}_\rho\nu_j+\text{hadron(s)})\right]
\end{align}
for energetically allowed $\rho$'s. We see (21) leads also to (20). 
    
Lastly we give a remark on the relations derived by assuming the 3 flavor number and 
$P(\pi^+(p)\rightarrow \mu^+\nu_j)$'s to be almost independent of all three $j$'s as well as by noting         
$m_\tau\simeq 1780$ MeV. Then, from (18) we obtain 
\begin{align}
&\ev{N_{\nu_e};\pi^+(p),\infty}\ :\ \ev{N_{\nu_\mu};\pi^+(p),\infty}\ :\ \ev{N_{\nu_\tau};\pi^+(p),\infty}\nn\\
&\simeq \sum_j|Z^{1/2}_{e j}|^2|Z^{1/2}_{\mu j}|^2\ :\ \sum_j|Z^{1/2}_{\mu
 j}|^4\ :\ \sum_j|Z^{1/2}_{\tau j}|^2|Z^{1/2}_{\mu j}|^2.
\end{align}
Similarly in the case of $K^+$, under the $j$-independence of $P\big(K^+\rightarrow
\bar{l}_\sigma \nu_j(\text{+ hadron(s)})\big)$
for $\sigma=e,\mu$, we obtain 
\begin{align}
&\ev{N_{\nu_e};K^+(p),\infty}\ :\ \ev{N_{\nu_\mu};K^+(p),\infty}\ :\ \ev{N_{\nu_\tau};K^+(p),\infty}\nn\\
&\simeq \sum_j|Z^{1/2}_{e j}|^2 R_j\ :\ \sum_j|Z^{1/2}_{\mu j}|^2 R_j\ :\
 \sum_j|Z^{1/2}_{\tau j}|^2 R_j 
\end{align}
with 
$R_j\equiv |Z^{1/2}_{e j}|^2 r(K^+\rightarrow e^+)+|Z^{1/2}_{\mu j}|^2 r(K^+\rightarrow \mu^+)$;
here, $r(K^+\rightarrow \bar{l}_\sigma)$ means the branching ratio of $K^+$ decay mode
with one $\bar{l}_\sigma$. If it is 
allowed for us to drop $r(K^+\rightarrow e^+)$ owing to the experimental value 
$r(K^+\rightarrow e^+)/r(K^+\rightarrow \mu^+)\simeq 4.87/66.70 \simeq 1/13.7$, 
we see RHS of (23) is nearly equal to that of (22). 

As to the neutron life time, we also obtain 
\begin{align}
\left[\frac{1}{2}\sum_r\ev{N_\rho(x^0);n(p,x^0_I)}_{con}/T\right]_{T\rightarrow
 \infty}
={\sum_j}'|Z^{1/2}_{\rho j}|^2 |Z^{1/2}_{e j}|^2 P(n(\vec{p})\rightarrow pe^-\bar{\nu}_j).
\end{align}
Assuming the independence of $P(n\rightarrow p\ e\ \bar{\nu}_j)$ on $j$'s, we obtain 
\begin{align}
&\ev{N_{\nu_e};n(p),\infty}\ :\ \ev{N_{\nu_\mu};n(p),\infty}\ :\ \ev{N_{\nu_\tau};n(p),\infty}\nn\\
&\simeq \sum_j|Z^{1/2}_{e j}|^4\ :\ \sum_j|Z^{1/2}_{\mu j}|^2|Z^{1/2}_{e j}|^2\
 :\ \sum_j|Z^{1/2}_{\tau j}|^2|Z^{1/2}_{e j}|^2.
\end{align}

In the standard theory of the neutrino oscillation in vacuum\cite{rf:6}, the transition 
probability for $\nu_\sigma\rightarrow \nu_\rho$ is given by  
\begin{align}
P_{\nu_\sigma \rightarrow \nu_\rho}(t=x^0-x^0_I)=|\sum_j U_{\sigma j}
 e^{i\omega_j(\vec{k})t} U_{\rho j}^\ast|^2,\quad U^\dagger U=I.
\end{align}
The time average of $P_{\nu_\sigma\rightarrow\nu_\rho}(t)$\cite{rf:6,rf:7} over 
a sufficiently long time interval $T\gg 4\pi k/|\Delta m^2_{ji}|$ for any pair
$(j,i),\ j\neq i$ is given by 
\begin{align}
\ev{P_{\nu_\sigma \rightarrow \nu_\rho}}_T \equiv \int^T_0 dt P_{\nu_\sigma
 \rightarrow \nu_\rho}(t)/T =\sum_j |U_{\rho j}|^2 |U_{\sigma j}|^2, 
\end{align}
which leads to the same ratio relations as (22) and (25) if we set $U=Z^{1/2}$.  Although 
this equality is taken in the standard theory,\cite{rf:6,rf:7} there is some problem for
understanding this equality on the basis of the quantum field theory, as first stressed by Blasone and  
Vitiello; a related remark is to be given below \citen{rf:4,rf:5,rf:11}.
From our viewpoint, the relation (27) should be understood as the 
relations (22) and (25) which have a field theoretical basis. 

\noindent {\it Final remarks and conclusion} --- 
It seems meaningful for us to consider whether or not we can construct such one $\nu_\rho$-state
(with momentum-helicity $(\vec{k},r)$) that the absolute square of the transition matrix element 
of $A^+ \rightarrow \bar{l}_\sigma \nu_\rho$ leads to Eq.(15).

We define
\begin{align}
\ket{\Psi_{\nu_\rho}(\vec{k}r; x^0)^d} \equiv 
 \frac{1}{\sqrt{V}}\int\!\!d^3x e^{i\vec{k}\cdot\vec{x}}(\nu_\rho(x)P(r))^d\ket{0},
\end{align}
where $P(r)$ is the helicity projection matrix\cite{rf:3}\tocite{rf:5}; ``$d$'' specifies the 
4-spinor component. RHS of (28) is expressed as 
$\sum_j {Z^{1/2}}^\ast_{\rho j} \alpha^\dagger_j(kr;x^0) u_j^\dagger(kr)^d \ket{0}$ with 
$\alpha^\dagger_j(kr;x^0)=\alpha^\dagger_j(kr)\exp(i\omega_j(k)x^0)$ ; thus, we have 
$\sum_d\bracket{\Psi_{\nu_\sigma}(\vec{k}r;x^0)^d}{\Psi_{\nu_\rho}(\vec{k}'s;x^0)^d}
=\delta_{\sigma\rho}\delta(\vec{k},\vec{k}')\delta_{rs}$. Let us consider the transition matrix
element 
\begin{align}
M(A^\dagger(p) \rightarrow \bar{l}_\sigma(qs)\nu_\rho(kr);x^0,x^0_i)^d
 &\equiv \bra{\Psi_{\nu_\rho}(\vec{k}r;x^0)^d}\beta_{l_\sigma}(qs;x^0) \nn\\
 &\times (-i)\int^{x^0}_{x^0_I}\!\!d^4z H_{\text{int}}(z) \alpha_A^\dagger(p;x^0)\ket{0},
\end{align}
where $\beta_{l_\sigma}(qs;x^0)=\beta_{l_\sigma}(qs)\exp(iE_\sigma(q)x^0)$, 
$\alpha_A(p;x^0)=\alpha_A(p)\exp(iE_A(p)x^0)$. As easily confirmed, we obtain 
\begin{align}
\sum_{\vec{q},\vec{k}}\sum_{\sigma,d,r,s}
 \bigg|M(A^\dagger(p) \rightarrow \bar{l}_\sigma(qs)\nu_\rho(kr);x^0,x^0_i)^d\bigg|
 = \text{RHS of (15)}.
\end{align}
This relation suggests that the state (28) is an appropriate one $\nu_\rho$ state (with
$(\vec{k},r)$) in quantum field theory. If so, we have 
\begin{align}
\sum_{d}\bracket{\Psi_{\nu_\rho}(\vec{k}r;x^0)^d}{\Psi_{\nu_\sigma}(\vec{k}r;x^0_I)^d}
 = \sum_j Z^{1/2}_{\rho j} e^{-i\omega_j(x^0-x^0_I)}{Z^{1/2}_{\sigma j}}^\ast,
\end{align}
leading to (26) with $U=Z^{1/2}$.

A simple way for obtaining the oscillation in space is to substitute $L$ (the distance) for $t$ and
to employ the wave-packets of the relevant particles. Along this line, it is necessary for us to
make clear what differences in the oscillation behavior exist among the present approach, the
standard formula and the approach of intermediate virtual neutrinos.\cite{rf:12,rf:13}

The authors would like to express their thanks to Prof.K.Ishikawa and other members 
of the particle theory group in Hokkaido University for continuous interest and discussions.   
Thanks are also due to Prof.G.Vitiello, Drs.M.Blasone and C.Giunti for their helpful 
discussions and kind hospitality.

\end{document}